\documentclass[useAMS,usenatbib,times,letter,amssymb]{mn2e}
\usepackage{epsfig,times,amssymb,amsmath,color,verbatim}
 
\usepackage{amsmath}
\usepackage{hyperref}
\usepackage{graphicx}
\usepackage{dcolumn}
\usepackage{bm}
\usepackage{natbib}
\usepackage{tabularx}

\def\ltsima{$\; \buildrel < \over \sim \;$}
\def\simlt{\lower.5ex\hbox{\ltsima}}
\def\gtsima{$\; \buildrel > \over \sim \;$}
\def\simgt{\lower.5ex\hbox{\gtsima}}

\def\[{\begin{equation}}
\def\]{\end{equation}}
\def\m@th{\mathsurround=0pt }
\def\eqalign#1{\null\,\vcenter{\openup1\jot \m@th
 \ialign{\strut\hfil$\displaystyle{##}$&$\displaystyle{{}##}$\hfil
 \crcr#1\crcr}}\,}

\begin{document}
\title{Enhancing the Cosmic Shear Power Spectrum}

\author[Simpson et al.]{Fergus Simpson,$^{1}$\thanks{email: fergus2@icc.ub.edu} 
Joachim Harnois-D\'eraps,$^{2}$
Catherine Heymans,$^{3}$  
Raul Jimenez,$^{1,4,5}$ 
\newauthor Benjamin Joachimi$^{6}$ 
and Licia Verde$^{1,4, 7}$\\
$^1$  Instituto de Ciencias del Cosmos, University of Barcelona, UB-IEEC, Marti i Franques 1, E08028, Barcelona, Spain.\\
$^2$ Department  of  Physics  and  Astronomy,  The  University  of  British  Columbia,
6224  Agricultural  Road,  Vancouver,  B.C.,  V6T  1Z1,  Canada. \\
$^3$ Institute for Astronomy, University of Edinburgh, Royal Observatory, Blackford Hill, Edinburgh EH9 3HJ, UK.\\
$^4$ ICREA (Instituci\'o Catalana de Recerca i Estudis Avan\c{c}at).\\
$^5$ Institute for Applied Computational Science, Harvard University, MA 02138, USA.\\
$^6$ Department of Physics and Astronomy, University College London, Gower Street, London WC1E 6BT, UK.\\
$^7$ Institute of Theoretical Astrophysics, University of Oslo, Oslo 0315, Norway.}

\date{\today}
\newcommand{\ud}{\mathrm{d}}
\newcommand{\fpe}{f_\perp}
\newcommand{\fpa}{f_\parallel}
\newcommand{\om}{\Omega_{\rm m}}
\newcommand{\dmax}{{\delta^{\rm{max}}}}
\newcommand{\kmax}{{k_{\rm{max}}}}
\newcommand{\Veff}{{V_{\rm{eff}}}}
\newcommand{\eff}{{\rm{eff}}}
\newcommand{\lcdm}{$\Lambda$CDM }
\newcommand{\hmpc}{ \, h \rm{ Mpc}^{-1}}
\newcommand{\hmpcv}{ \, h^3 \rm{ Mpc}^{-3}}
\newcommand{\hinvmpc}{ \, h^{-1} \rm{ Mpc}}
\newcommand{\tripleint}{\int \! \! \int \! \! \int}
\newcommand{\kfund}{k_f}
\newcommand{\dint}{\int  \!\!  \int}
\newcommand{\dclip}{\delta_{c0}}
\newcommand{\dThresh}{\delta_{0}}
\newcommand{\deltaX}{\delta^{(X)}}
\newcommand{\rh}{{\rho}(r)}
\newcommand{\clip}{{\mathcal{C}}}
\newcommand{\Plin}{{P_\mathrm{L}(k)}}
\newcommand{\Ponetwo}{{P_{\mathrm{c}12}(k)}}
\newcommand{\doa}{{ \delta_{0}}} 
\newcommand{\fid}{{\mathrm{fid}}}
\newcommand{\amp}{{ \alpha }}
\newcommand{\fsig}{ f \sigma_8}
\newcommand{\Pshear}{P_\kappa}
\newcommand{\persqrArcmin}{\mathrm{ arcmin}^{-2}}
\newcommand{\Cl}{P_\kappa}    
\newcommand{\surveyarea}{{$10$,$000$}}

 \newcommand{\xk}{\bmath{\theta}}
 \newcommand{\xvec}{\bmath{x}}

 \newcommand{\omseight}{\om - \sigma_8}

\newcommand{\scalar}{\delta}
\newcommand{\threshold}{{\scalar_0}}
\newcommand{\clipfield}{\scalar_c}
\newcommand{\obsfield}{\scalar_n} 

\newcommand{\Fisher}{\mathrm{F}}

\date{\today}

\maketitle

\begin{abstract}
Applying a transformation to a non-Gaussian field can enhance the information content of the resulting power spectrum, by reducing the correlations between Fourier modes.  In the context of weak gravitational lensing, it has been shown that this gain in information content is significantly compromised by the presence of shape noise. We apply clipping to mock convergence fields, a technique which is known to be robust in the presence of noise and has been successfully applied to galaxy number density fields. When analysed in isolation the resulting convergence power spectrum returns degraded constraints on cosmological parameters. However substantial gains can be achieved by performing a combined analysis of the power spectra derived from both the original and transformed fields. Even in the presence of realistic levels of shape noise, we demonstrate that this approach is capable of reducing the area of likelihood contours within the $\om - \sigma_8$ plane by more than a factor of three. 
\end{abstract}

\section{Introduction}
 
The extraction of useful information from cosmological fields (such as galaxy number densities or weak lensing) is often heavily compromised by two limiting factors. Firstly the smallest physical scales are too challenging to model theoretically and are therefore excluded from the analysis. This is particularly problematic for more evolved fields where the amplitude of fluctuations are sufficiently large that the particle dynamics behave in a highly non-linear manner. Secondly, for any non-Gaussian field, the power spectrum fails to retain all of the statistical information captured by the field's configuration. The amplitude of the field's Fourier modes are correlated, thereby compromising the information content of the power spectrum. These correlations can be interpreted as originating from the largest dark matter halos, which form spikes in the density field. Each spike in real space translates to a uniform signal in Fourier space, making a coherent contribution to the power spectrum across all wavelengths.  Both of these issues, the modelling of small scale clustering and the correlation of Fourier modes, are predominantly associated with the highest density regions of the field. 
 
Manipulating the observed field prior to evaluating the power spectrum can help alleviate these problems. Suppressing fluctuations within the highest density regions in the field improves the performance of perturbation theory, allowing a greater number of Fourier modes to be included in a likelihood analysis.  This has been demonstrated with simulated galaxy fields \citep{2013SimpsonClip} and applied to data from the Galaxy And Mass Assembly (GAMA) galaxy redshift survey \citep{2015SimpsonGAMA}.  The suppression of non-Gaussian peaks also leads to a reduction in the correlation between Fourier modes, thereby enhancing the amount of information retained by the power spectrum. While previous applications of clipping involved three-dimensional fields, in this work we turn our attention to weak gravitational lensing which involves a projection of the cosmological density field onto two dimensions.  

The apparent alignment of galaxies on the sky arises from the optical distortion imposed by the intervening distribution of inhomogeneous matter.  When considering the two-point statistics associated with this weak gravitational lensing signal, a prominent degeneracy emerges between two cosmological parameters: the amplitude of linear density perturbations $\sigma_8$ and the cosmic matter density $\om$.  For example, the cosmic shear correlation function from the Canada-France-Hawaii Telescope Lensing Survey (CFHTLenS) has been found to measure the combination $ \sigma_8 \left( \om / 0.27 \right)^{0.6} = 0.79 \pm 0.03$ within the context of a standard flat \lcdm cosmology \citep{Kilbinger2012}. There are a number of possible routes to improve this measurement. Using either tomography \citep{syspaper} or the full three-dimensional information \citep{2014Kitching3D} could lead to significant gains.  Alternatively, further information can be extracted from the field by analysing the three-point statistics, as demonstrated by \citet{2014FuCFHTLenS}, or the abundance of peaks \citep{2014CardonePeaks, 2002ApJ...575..640W, 2010ApJ...712..992B, 2010A&A...519A..23M, 2010Dietrich}. 

\citet{2011ApJ...729L..11S} found that by invoking a logarithmic transformation, considerable improvements could be made to the information content of the convergence power spectrum, while \citet{2012MNRAS.421..832Y}  demonstrated that further enhancements were achievable by performing a wavelet decomposition of the convergence field.  \citet{2011MNRAS.418..145J} present a comprehensive analysis of applying Box-Cox transformations (among which the logarithmic transforms form a subset) to the convergence field. They also reported that significant enhancements to the power spectra were achievable under idealised conditions. However each study concluded that realistic levels of shape noise substantially reduce these gains. 

Our work aims to build on these previous studies in two ways. First of all we shall focus on the clipping transformation which is well suited to noisy fields. Secondly, we will investigate the potential benefits of combining the information from both spectra, before and after the transformation, by taking into account their cross-covariance. 

In \S \ref{sec:clip} we briefly review how local transformations may be applied to weak lensing fields. Methods for estimating the covariance matrices via numerical simulations are detailed in \S \ref{sec:methods}, followed by our main results which are presented in \S \ref{sec:results}. Our conclusions are summarised in \S \ref{sec:conclusions}.

\section{Transforming Weak Lensing Fields} \label{sec:clip}

Clipping is a local transformation characterised by the application of a saturation value $\threshold$ to a scalar field $\scalar(\xvec)$ such that

\[ \label{eq:clip}
\eqalign{
\clipfield (\xvec) &= \threshold \qquad (\scalar (\xvec) > \threshold )      \cr
\clipfield(\xvec)  &=  \scalar(\xvec) \quad   (\scalar (\xvec) \leq  \threshold )  \cr
}
\]
\noindent  yielding the clipped field $\clipfield(\xvec)$. The motivation for applying clipping to cosmological fields is twofold: (1) the potential improvement in modelling smaller scale clustering, and (2) the introduction of new information to the power spectrum. In this work we shall focus on the latter.

This transformation has the unique property of either eliminating or unperturbing the existing noise field. Consider a noise field $n$, defined as the difference between the original field $\scalar (\xvec)$ and observed field $\obsfield (\xvec)$
\[
n(\xvec) =  \obsfield(\xvec) - \scalar(\xvec) \, ,
\]
then the transformed noise field $n'(\xvec)$ is the difference in the transformed fields
\[
\eqalign{
n'(\xvec) &= f(\obsfield) - f(\scalar) \, , \cr
&= n(\xvec) f'(\scalar) \, ,
}
\]
where $f'$ is the functional derivative of $f$. For clipping, as specified by (\ref{eq:clip}), $f'(\scalar)$ can only take two values:  zero or unity. Therefore, the noise contribution is either removed (in regions above the threshold) or left untouched (in regions below the threshold).

The benefits of decorrelating the Fourier modes within the galaxy power spectrum are limited since its amplitude is dictated by the uncertain manner in which galaxies trace the underlying matter distribution. Conversely, enhancing the weak lensing power spectrum would be of greater consequence as it acts as a direct measure of the amplitude of matter fluctuations.

A possible disadvantage of applying clipping to lensing stems from the fact that lensing involves a projection of the three dimensional density field onto two dimensions. This averaging process will reduce the level of non-Gaussianity in the field, and so a greater proportion of the field must be subject to clipping in order to achieve the desired result. 
 
While the direct observable from weak lensing surveys are the shapes of galaxies, it is the derived convergence field which directly relates to the projected mass density. The convergence field $\kappa(\xk)$ is therefore the more natural quantity to work with. It may be thought of as a weighted integral of the matter perturbations along the line of sight

\[\label{eq:kappa}
\kappa(\xk) \simeq \frac{3}{2} \left(\frac{H_0}{c} \right)^2 \om \int g(\chi) f_K(\chi) \frac{\delta(\xk, \chi) }{a(\chi)}  \ud \chi \, ,
\]
where  $\chi$ denotes the radial coordinate distance.  The lensing efficiency $g(\chi)$ is determined by the radial distribution of source galaxies $n_g(\chi)$ and the comoving angular diameter distance $f_K(\chi)$

\[ \label{eq:lensefficiency}
g(\chi) = \int_\chi^{\infty} \ud \chi' n_g(\chi') \frac{f_K(\chi'-\chi)}{f_K(\chi')}  \, .
\]
\noindent If we decompose the matter field into linear and non-linear components, $\delta  = \delta_G + \delta_x$, where $\delta_G$ is the underlying Gaussian random field and the residual term $\delta_x$ incorporates all departures from Gaussianity, then from Eq.~(\ref{eq:kappa}) the convergence is separable in the same manner, yielding $\kappa(\xk) = \kappa_G(\xk) + \kappa_x(\xk)$. In the limit that the non-Gaussian component $\kappa_x(\xk)$ only contributes at field values larger than the chosen threshold, the application of clipping as defined by Eq.~(\ref{eq:clip})  ensures that only the Gaussian component of the density field $\delta_G(\xk)$ contributes to the resulting convergence power spectrum $\Cl$. 

For comparison we shall also assess the performance of the logarithmic transformation \citep{2011ApJ...729L..11S, 2011MNRAS.418..145J}

\[  \label{eq:log}
 \bar{\kappa}(\xk)  = \kappa_0  \ln \left[ 1 + \kappa (\xk)/\kappa_0 \right] \, ,
\]
where the parameter $\kappa_0 $ can be adjusted to compensate for the amplitude of the convergence field, serving a similar purpose to the clipping threshold $\threshold$. We follow the prescription used in \citet{2011ApJ...729L..11S} and set $\kappa_0 = |\kappa_{\mathrm{min}} | + 0.001$, where $\kappa_{\mathrm{min}}$ is the most negative value of the convergence. This small offset ensures that the transformed field remains finite.

\section{Methodology} \label{sec:methods}

In order to quantify the benefits of clipping the convergence field, we perform a Fisher matrix analysis to determine the error forecast in the $\omseight$ plane for a fiducial survey, and compare the size of the expected likelihood contours before and after the convergence field is subject to a local transformation.

 It is well known that in circumstances where the posterior is not well described by a multivariate Gaussian, the Fisher matrix acts as a poor estimator (see for example \citet{2014MNRAS.441.1831S, 2015arXiv150605356S}). However when considering forecasts for weak lensing surveys in the $\omseight$ plane, then provided the likelihood contours are sufficiently small, comparisons between Fisher predictions and Markov Chain Monte Carlo (MCMC) outputs are found to be highly consistent (see for example Figure 3 of \citealt{2010MNRAS.408..865T} and Figure 6 of \citealt{2012Wolz}).

\subsection{Simulations}

We make use of mock convergence fields from simulations described in \citet{2014JoachimSims}. We apply a variety of clipping thresholds to the mock convergence fields associated with 70 independent lines of sight, each of which spans 60 square degrees.  Sources are placed at a redshift of $z = 0.8$ and the fiducial amplitude of linear density fluctuations is taken to be $\sigma_8 = 0.826$.  These lines of sight serve as our estimator for the covariance of the angular power spectra $\Cl(\ell)$. In addition to these, we utilise a set of five simulations which are seeded with an initial particle distribution with matching phases, but with small displacements in the values of $\om$ and $\sigma_8$. This allows for a clean extraction of the partial derivative of the observables with respect to these parameters, as required for estimating the Fisher Matrices, without being contaminated by cosmic variance.  Other cosmological parameters such as the spectral index $n_s$ and global curvature $\Omega_k$ are taken to be fixed, as they can be measured with a much greater precision from other observations such as the Cosmic Microwave Background. 

For each line of sight the convergence field is assigned to a $300 \times 300$ grid, such that each grid cell corresponds to approximately 1.5 arcminutes, and the convergence  power spectra are evaluated in eight equally spaced bins for multipoles $\ell < 1,500$. Our results are largely insensitive to the choice of smoothing length. However if very fine grids are used then clipping predominantly acts on noise spikes rather than true peaks in the projected density field. 

When generating each simulated convergence field, the source redshift plane ought to remain fixed. However the ray-tracing takes place across an integer number of simulation boxes, and since the redshift-distance relation is altered when perturbing $\om$, there is inevitably a small unwanted shift in the redshift of the sources.We compensate for this by rescaling the amplitude of the convergence field, which accounts for the change in lensing efficiency $g(\chi)$ due to the small difference between the desired and simulated distances to the source plane. The magnitude of this correction is approximately $1\%$ and so would not change our results by more than a few percent if it were neglected. 

\subsection{Shape Noise}

Starting from the true convergence field, we use the inverted Kaiser Squires method \citep{1993KaiserSquires} to recover the shear field at each cell location. Then noise is superposed by drawing two random numbers at each cell location, from a Gaussian distribution with mean of zero and standard deviation $\bar{\sigma}_e / \sqrt{2}$, which constitute the real and imaginary parts of the mean intrinsic ellipticity of the galaxies within that cell. The cell averaged ellipticity  $\bar{\sigma}_e$ is given by ${\sigma}_e / \sqrt{n}$ where n is the number of source galaxies per cell, and we take the rms intrinsic ellipticity $\sigma_e = 0.3$ throughout this work. The noisy shear field is then given as a sum of the true and intrinsic shear fields. The noisy convergence field is recovered by applying the Kaiser-Squires transformation \citep{1993KaiserSquires} to the noisy shear field.

The intrinsic shape noise introduces an additive white noise contribution which includes a distortion associated with the window function of the grid

\[
P_n(k) = \frac{\sigma_e^2}{2 \bar{n}} \left< \frac{1}{\textrm{sinc}(k_x/2)^2} \frac{1}{\textrm{sinc}(k_y/2)^2} \right> \, ,
\]
\noindent where $k_x$ and $k_y$ specify the elements of the wavevector in the $x$ and $y$ dimensions, in units of the Nyquist frequency, and the averaging takes place within the annulus associated with the wavevector bin. Clipping reduces the noise contribution to the power spectrum, since the regions of the field lying above the threshold are smoothed. This leads to an estimate of the clipped shot noise contribution $P_{cn}(k)$, derived from equation (5) of \citet{2013SimpsonClip}, where the leading order term may be expressed as

\[  \label{eq:clipShotNoise}
P_{cn}(k) \simeq (1 - f_c)^2 P_n(k) \, ,
\]

\noindent where $f_c$ is the proportion of the field which lies above the clipping threshold.  For the fields considered in this work, the value of $f_c$ required to reduce the large scale power by a factor of two is around $20\%$. By comparison, the proportion of a three-dimensional galaxy field subject to a comparable clipping strength is typically $2\%$ \citep{2015SimpsonGAMA}.

\begin{figure}
\includegraphics[width=80mm]{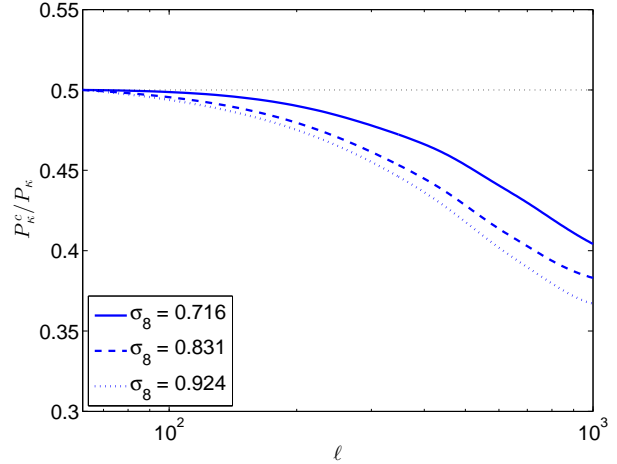}
\caption{The ratio of the convergence power spectrum from the clipped convergence field, to that of the original field, for three different values of $\sigma_8$.  In this example a clipping threshold is chosen such that on the largest angular scales the power is reduced by a factor of two, which typically provides a good balance between the removal of non-linear power while maintaining a high level of signal to noise. 
\label{fig:s8Ratios}}
\end{figure}
 
 \begin{figure*}
\includegraphics[width=80mm]{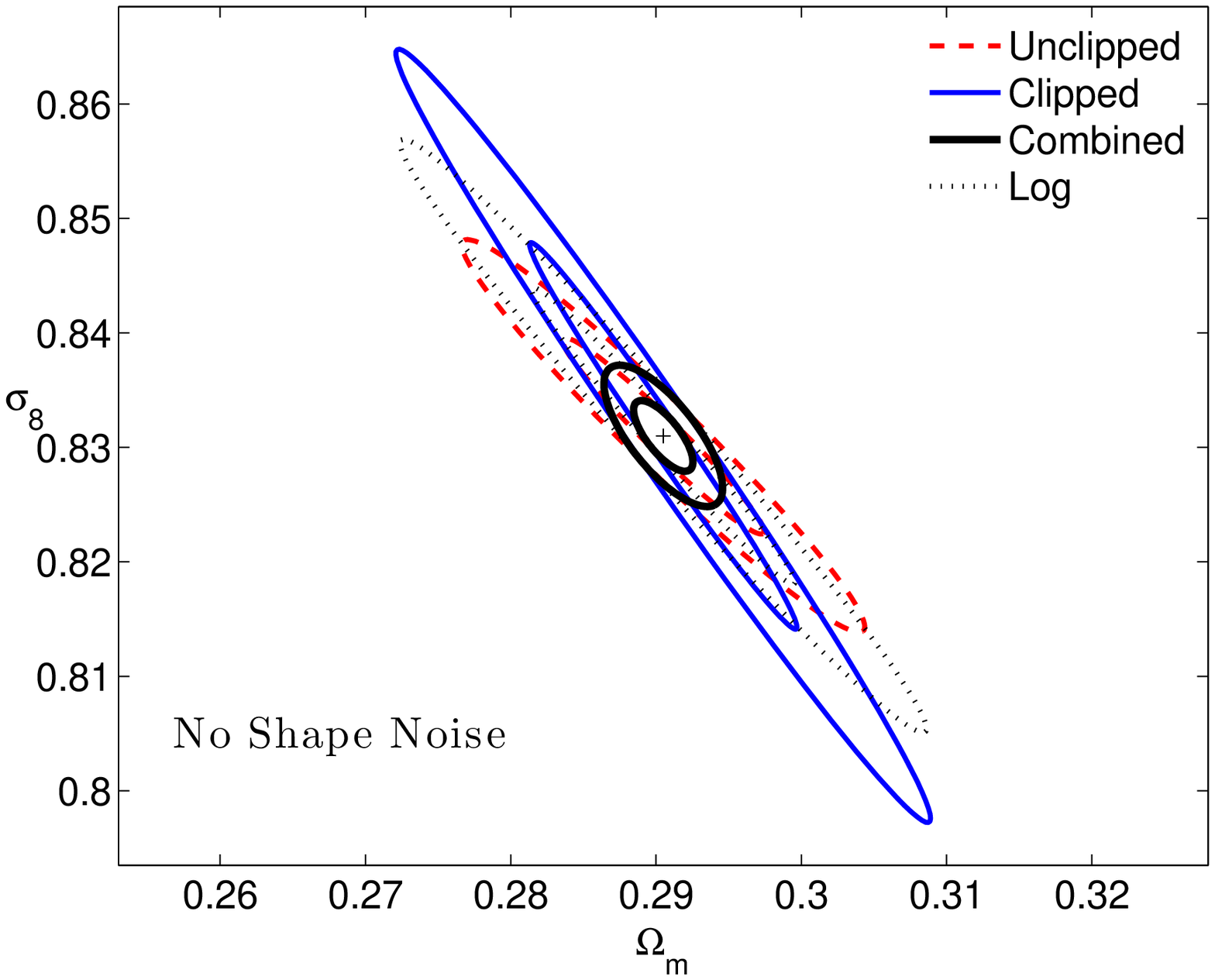}
\includegraphics[width=80mm]{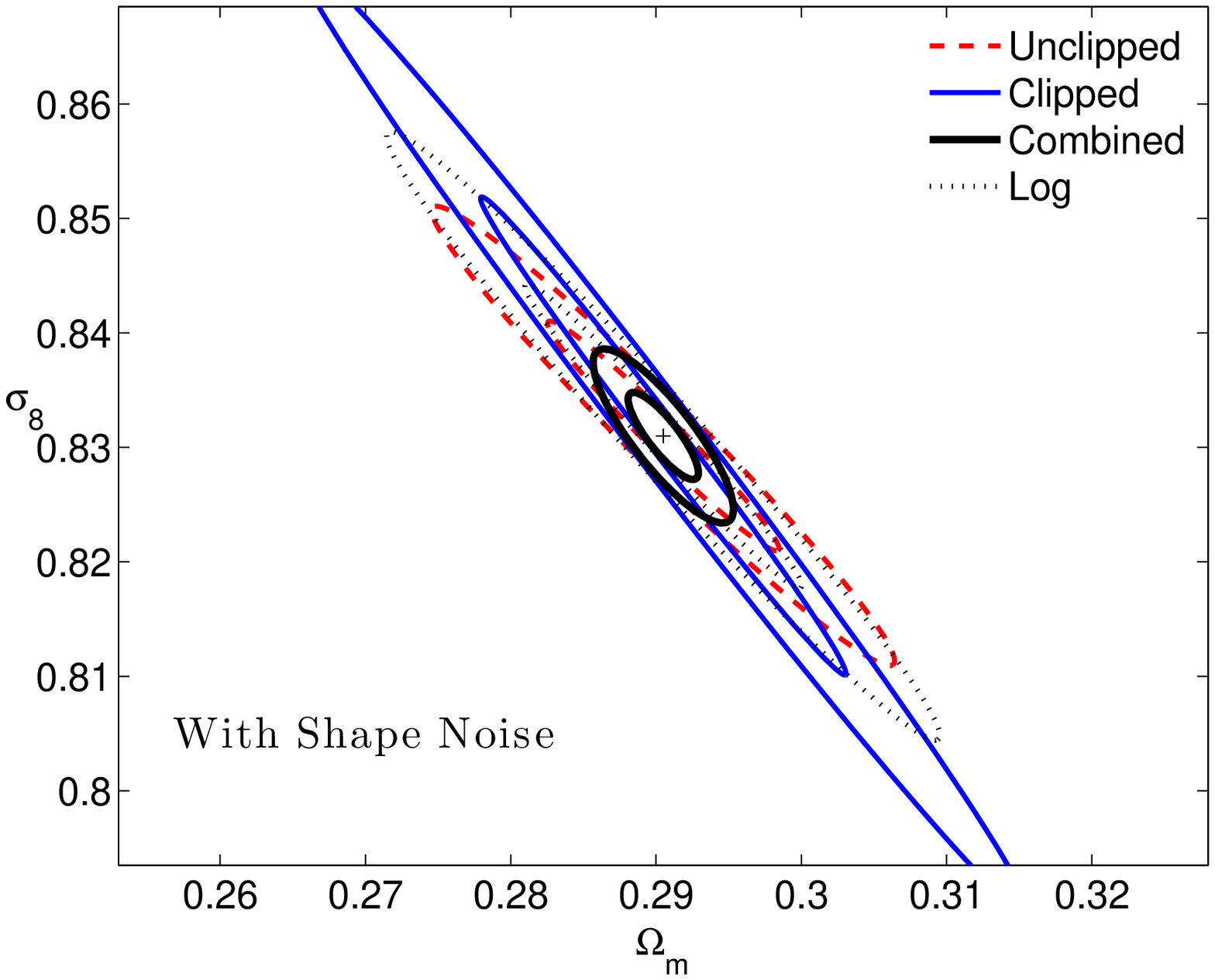}
\caption{\textit{Left:} Constraints in the $\omseight$ plane using the angular power spectra $(\ell < 1$,$500)$ of the original convergence field (dashed),   the clipped field (thin solid) and the log field (dotted). The thick set of contours relate to the combined analysis of power spectra from the original and clipped fields.  These constraints correspond to a survey of \surveyarea \, square degrees. \textit{Right:} The same format as the left panel, but we now include the effects of shape noise, with a galaxy number density of $20 \, \mathrm{arcmin}^{-2}$ and an rms intrinsic ellipticity $\sigma_e = 0.3$.     \label{fig:clipFisherPanels}}  
\end{figure*}

\subsection{Covariance Estimation} \label{sec:cov}

Since estimating the covariance matrix is a critical step in this analysis, and one which can be highly susceptible to the influence of noise, we compare two different methods for estimating the covariance matrix.  First we use the Ledoit-Wolf shrinkage estimator \citep{2004LedoitWolf} to evaluate the covariance matrix $\mathrm{\hat{C}}$  associated with our binned power spectra.
\[ \label{eq:cov}
\mathrm{\hat{C}} = \delta^{\star} {\mathrm F}  + (1 -  \delta^{\star} ) {\mathrm S} \, ,
\]
We use publicly available code\footnote{\protect\url{http://www.econ.uzh.ch/faculty/wolf/publications/covCor.m.zip}} to estimate the shrinkage target F and the shrinkage constant $\delta^{\star}$.  The form of the shrinkage target is based upon the samples being independent and identically distributed.  The sample covariance matrix S is defined as the ensemble average over $70$ lines of sight
\[
\mathrm{S}_{ij} = \langle \Delta \Cl^m(i) \Delta \Cl^n(j) \rangle \, ,
\]
where $\Delta \Cl^m(i)$ represents the deviation in power from the sample mean, and $m$ denotes the particular transformation(s) under consideration. The total number of columns in the covariance matrix is given by the product of the number of angular bins in $\Cl (\ell)$ with the number of transformations included in the analysis, in order to accommodate their cross-covariance.

The finite number of fields is a source of sampling noise which will inevitably propagate through to the estimation of our likelihood contours. In order to confirm that the uncertainty in our likelihood contours is negligible, we employ a second algorithm to estimate the covariance matrix. Instead of using a shrinkage estimator, we simply invert the sample covariance matrix S before applying a rescaling factor which accounts for the noise-induced bias in the amplitude  \citep{2007A&A...464..399H}

\[ \label{eq:rescale}
\mathrm{\hat{C}}^{-1} = \frac{n - p  - 2}{n - 1} \mathrm{S}^{-1} \, ,
\]
where $p$ represents the length of the data vector (either 8 or 16),  and $n$ is the number of samples. When all 70 fields are used to construct the sample covariance S,  these two algorithms produce highly  consistent results, the derived likelihood contours differ in area by less than $5\%$.
 
A further validation test for convergence of the covariance matrix involves repeating this procedure with a small subset of the available samples. This also proves useful in order to identify which algorithm produces more consistent results. 
For example, when using only 20 samples the rescaling method defined by Eq.~(\ref{eq:rescale}) systematically overestimates the area of the likelihood contour by more than $40\%$. By contrast, the shrinkage estimator recovers likelihood contours which are consistently within $40\%$ of the area derived from the full sample. This improved stability leads us to conclude that the shrinkage algorithm likely offers a superior performance. It also predicts slightly more conservative gains when combining the clipped and unclipped spectra. For these two reasons, the shrinkage algorithm defined by Eq.~(\ref{eq:cov}) is the one we adopt when presenting our main results.

\begin{table*}
\centering
\begin{tabular}{|l|c|c|c|c|c|}
\hline
&\textbf{Unclipped}&\textbf{Logarithm}&\textbf{Clip 40\%}&\textbf{Clip 60\%}&\textbf{Clip 80\%}\\\hline
\textbf{Unclipped}&1&1.4&3.1&3.3&2.9\\\hline
\textbf{Logarithm}         & &0.73&2.6&2.5&1.8\\\hline
\textbf{Clip 40\%}& & &0.59&0.87&1.5\\\hline
\textbf{Clip 60\%}& & & &0.59&1.2\\\hline
\textbf{Clip 80\%}& & & & &0.72\\\hline
 \end{tabular}
  \caption{The improvement in the Figure of Merit as defined in Eq.~(\ref{eq:FoM}), for various combinations of power spectra from transformed convergence fields, relative to that of the original power spectrum. The four local transformations - three clipping thresholds and the logarithmic transform -  all prove detrimental when considered alone, yet yield substantial gains when coupled with the power spectrum of the original field. }
  \label{tab:1}
\end{table*}

\subsection{Constructing the Fisher Matrix}

The precision with which a parameter may be determined is governed by the rate at which the observable responds to a change in the parameter. The key observable to consider is the power spectrum of the convergence field $\Pshear (\ell)$. This may be expressed as  

\[ \label{eq:shear}
\Pshear (\ell) = \frac{9}{4} \om^2 \left( \frac{H_0}{c} \right)^4 \int_0^{\infty} \frac{g^2(\chi)}{a^2(\chi)} P_\delta \left(\frac{\ell}{f_K(\chi)} , \chi \right) \ud \chi \, ,
\]
where the lensing efficiency $g(\chi)$ is defined in Eq.~(\ref{eq:lensefficiency}).

Predicting the power spectrum of a clipped convergence field poses a significantly more challenging problem.  As is apparent from Eq.~(\ref{eq:shear}), the convergence field can be thought of as arising from the weighted sum of intervening lens planes. Applying clipping to a convergence field derived from of a single lens plane would possess a power spectrum in line with the model presented in \citet{2013SimpsonClip}. However each of the stacked lens planes is at a slightly different stage of gravitational evolution. We therefore resort to using numerical simulations in order to determine the expected behaviour. In order to define a consistent clipping strength, independent of smoothing length, we specify the fractional drop in amplitude of the large scale power. We perform an iterative procedure on the threshold value $\threshold$ to converge upon the desired clipping strength. Note that this iteration is not performed for each field - it is performed once, then this threshold value is fixed for all remaining fields within the ensemble. Furthermore, prior to applying any transformation, each convergence field has its mean subtracted. This reflects the uncertain normalisation of a given convergence field derived from the measurement of a shear field, although this is less problematic for fields spanning a large area. 

Our fiducial survey spans  \surveyarea  \, square degrees, and involves a galaxy number density of $20 \, \mathrm{arcmin}^{-2}$. For simplicity we only consider a single redshift bin, rather than using tomography, and the mock survey is idealised in that there are no holes or masks. 

The Fisher information matrix, formally the expectation of the Hessian matrix of the log likelihood, may be expressed as

\[
\mathrm{F =  D^T {C}^{-1} D } \, ,
\]
where C is the covariance matrix corresponding to the various bins in $\Cl(\ell)$ and is estimated following the methodology in \S \ref{sec:cov}. Where a combined analysis of two spectra is performed, the covariance matrix incorporates the cross-covariance between the two spectra. The Jacobian matrix D is comprised of elements $\mathrm{D}_{ij}$ which hold the differential of the $i$th multipole bin with respect to the $j$th parameter,

\[
\mathrm{D}_{ij} = \frac{\partial \Cl(i)}{\partial p_j} \, .
\]
We expect transformed fields will be susceptible to shape noise and so D must be re-evaluated for each transformation when noise is introduced, as discussed in further detail in \citet{Seo2012}. We take the average power derived from twenty different noise realisations in order to ensure that the above derivatives are stable.  

There are two distinct analyses we wish to perform in the $\omseight$ plane. The first is the constraint which one would achieve from a conventional interpretation of the weak lensing power spectrum, without clipping. In this case the determination of the differential matrix D can be performed analytically. This is not the case for the power spectra derived from the clipped convergence fields. In this case, to determine the differential matrix D we utilise the simulations which were seeded with matching phases in the initial density fluctuations. Due to the nonlinear nature of structure formation, the smallest density fluctuations exhibit a stochastic behaviour across the different simulations, but on the angular scales of interest we find the derivatives to be numerically stable.

\section{Results from Simulations} \label{sec:results}

Using the simulations outlined in the previous section, we explore the consequences of applying local transformations to the convergence field, and construct likelihood contours derived from the resulting convergence power spectra. 

\subsection{Power Spectra}
  
Figure \ref{fig:s8Ratios} shows the change in the shape of the convergence power spectrum for three different cosmologies, after applying a clipping threshold chosen such that the largest angular bin in $C_\ell$ is reduced in amplitude by a factor of two. Previous studies of clipping cosmological fields conclude that reducing the linear power by between $40 - 60\%$ strikes a good balance by suppressing higher order terms, while maintaining a high level of signal to noise. This therefore drives our choice of the threshold value, which for the central value ($\sigma_8 = 0.831$) corresponds to $\kappa_0 = 0.026$.  The three different values of $\sigma_8$ generate significantly different responses, illustrating how clipping can reveal the degree of non-linearity present in the field. For more evolved density fields,  a greater suppression of power occurs on small angular scales. This response can be understood by considering the original field to be a superposition of its linear and nonlinear components. The clipped power is then predominantly comprised of the clipped linear power spectrum and the clipped nonlinear power \citep{2013SimpsonClip}. However owing to the positive skewness of the nonlinear pdf, the nonlinear power suffers a much stronger loss of power than the Gaussian pdf. Thus, fields which are more evolved and host a stronger nonlinear component suffer a greater loss of power on small scales. This behavior has previously been observed for the case of the three-dimensional simulated density fields  \citep{2013SimpsonClip}. Applying stronger clipping with a lower threshold value leads to a more linear power spectrum with less correlated Fourier modes. This improvement saturates once the amplitude of the linear power spectrum is reduced by approximately a factor of two, so there is little to be gained by adopting a lower threshold. 

\subsection{Parameter Constraints}

The left hand panel of Figure \ref{fig:clipFisherPanels} illustrates the $68$ and $95\%$ joint likelihood contours derived from the Fisher matrix of the convergence power spectrum $(\ell < 1500)$. The dashed lines are derived from the standard convergence field, while the dotted lines are derived from the log transform as defined by Eq.~(\ref{eq:log}). The outer set of solid contours make use of the clipped convergence field as defined by Eq.~(\ref{eq:clip}), with a threshold selected such that the large scale power is reduced to $50\%$ of its original value. These contours are somewhat elongated relative to the original spectrum, but also exhibit a small clockwise rotation of their degeneracy direction. The inner set of solid contours result from combining both the original and clipped power spectra.   The minor axes of the combined contours are not noticeably reduced from those of the original spectrum, owing to the strong cross-covariance between the amplitudes of the two spectra. However the major axes shrink considerably, suggestive of a powerful complementarity between the information held by the two power spectra.

Correlations between data points are often perceived to have a detrimental impact on the dataset's overall information content. But there are many circumstances in which this is not the case. To see why, consider the simple case of two data points, to which we wish to fit a straight line. The only two free parameters are the gradient and the amplitude. If the errors on these two data points are perfectly correlated, the constraint on the amplitude weakens by a factor of $\sqrt{2}$ relative to the uncorrelated case. However the gradientÔs uncertainty vanishes - it is now determined to an arbitrarily high precision. 

This toy model is intimately linked to the case we consider here. For weak lensing contours in the $\om-\sigma_8$ plane,  the size of the minor axis relates to the uncertainty in the amplitude of the weak lensing power spectrum. This is often quoted as a measurement of the parameter combination $\sigma_8 \om^\alpha$. The fact that the clipped and unclipped spectra are highly correlated means that when a joint analysis is performed, any improvement in the measurement of this amplitude is negligible. Hence the minor axis does not shrink for the case of the `Combined' contours in Figure 2. Meanwhile the extent of the contour's major axis is governed by how well the shape of the power spectrum is measured. More evolved fields with higher values of $\sigma_8$ will possess more power on smaller angular scales, and this helps lift the degeneracy within the $\om -  \sigma_8$ plane. When considered separately, the precision with which the shape of each spectra can be determined is ultimately limited by cosmic variance. It might be that the large scale power is low by chance, causing a spurious upward tilt in the lensing power spectrum, thereby mimicking a higher value of $\sigma_8$. However this deception can be revealed with a joint analysis of both spectra. A higher value of $\sigma_8$ leads to a stronger decrement of small scale power in the clipped spectrum relative to the unclipped. This relative change in power is largely insensitive to the particular cosmic variance realisation. The contour's major axis therefore shrinks further than would have been possible if the two spectra were uncorrelated. This is explored further in  \S \ref{sec:boxcox} and Figure \ref{fig:boxcox} below.

Note that this is not the only cosmological example where correlations prove to be beneficial. Performing an analysis of redshift space distortions with multiple tracers allow cosmic variance fluctuations to be effectively nulled, permitting an enhanced measurement of the growth of structure. 

These results are consistent with the findings of \citet{2011MNRAS.418..145J} who noted that upon performing a Box-Cox transformation the likelihood contours demonstrate a small clockwise rotation of the degeneracy direction, and become considerably elongated. This elongation of the contours for transformed fields occurs because information regarding the shape of the spectra (as induced by non-linear structure associated with higher values of $\sigma_8$) is erased. Indeed the full extent of this elongation effect is likely lost in our case due to the limitations of the Fisher formalism. Nonetheless the addition of the original information from the unclipped field truncates the longer contours. The combined contours therefore offer a faithful representation of the true likelihood. 

Our findings differ from \cite{Seo2012} in that even in the noise-free case, we do not observe significant gains from applying a logarithmic transform alone. However there are a number of differences in our methodologies. They use multiple redshift bins and explore a broader parameter space - for example we do not marginalise over the dark energy parameters $w_0$ and $w_a$. Furthermore the efficacy of the log transform is strongly governed by the choice of pixel size.

 The right hand panel of Figure \ref{fig:clipFisherPanels} takes the same format as the left panel, but here we include the impact of shape noise.  There is only a modest degradation of the unclipped contours in this case. The degeneracy direction of the transformed fields experiences a slightly reduced rotation, but there is still sufficient complementarity to yield substantial gains when performing a combined analysis. The logarithmic contours also show little degradation, which may appear surprising given the information loss found by \cite{2011ApJ...729L..11S}. However the impact of shape noise is highly sensitive to the choice of smoothing length used to define the field, particularly since the amplitude of the convergence fundamentally alters the transformation defined in (\ref{eq:log}), and here we are using a considerably coarser pixel size.
 
The extent to which the benefits of applying clipping and logarithmic transformations are degraded by the presence of shape noise also depends on the specific choice of transformation chosen. Higher clipping thresholds lead to a higher power spectrum and thus a lower susceptibility to shape noise. Yet they are also less efficient at recovering information to the power spectrum in the low noise limit. 

In order to quantify these gains we can readily compute the change in the figure of merit such as those previously defined in \citet{2009arXiv0901.0721A} and \citet{simpsonp09}
\[ \label{eq:FoM}
\mathrm{FoM} = \sqrt{\Fisher_{\Omega \Omega} \Fisher_{\sigma \sigma} - \Fisher^2_{\Omega \sigma}} \, ,
\]
where the notation $\Fisher_{\theta \theta}$ denotes the elements of the Fisher matrix associated with the parameter $\theta$. This quantity is inversely proportional to the area enclosed by the likelihood contours.  The absolute value is not of great interest, since it will strongly depend on the configuration and area of the survey in question, so in Table \ref{tab:1} we quote in each case the fractional change in the Figure of Merit relative to the original  power spectrum analysis. We show results for three different clipping thresholds, corresponding to those which reduce the large scale power to $40$, $60$ and $80\%$ of its original value. In addition we include the original convergence field, and the log transform as defined by Eq.~(\ref{eq:log}).  The diagonal elements demonstrate that for each transformation under consideration, the areas enclosed by the ellipses have enlarged. The off-diagonal elements in Table \ref{tab:1} quantify the significant gains which can be made by performing a combined analysis, particularly when combining the original spectrum with that from a clipped field, where the gain is more than a factor of three, despite the presence of shape noise. We have also evaluated intermediate clipping strengths, finding that the gain derived from combining two clipped power spectra deteriorates rapidly as the difference between their clipping strengths is reduced. 

\begin{figure}
\includegraphics[width=80mm, angle=270]{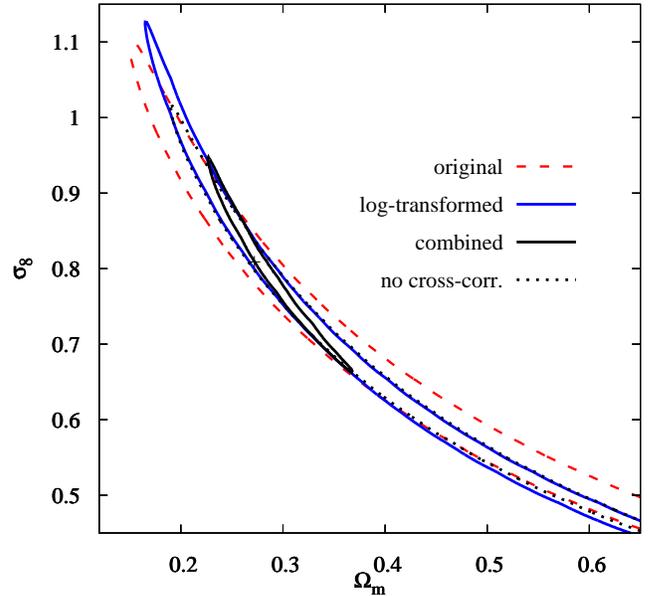}  
\caption{Simulated constraints for a 200 square degree convergence field, before (red dashed contours) and after (blue solid contours) applying a logarithmic transformation (95\% C.L.). The inner black solid contour represents a combined analysis of both spectra. The consequence of ignoring the cross-covariance between the original and transformed spectra is illustrated by the dotted black contour. Shape noise is not included in the convergence field. The black marker signifies the fiducial cosmology of the simulation.}
\label{fig:boxcox}
\end{figure}

Thus while the findings of previous related studies \citep{2011ApJ...729L..11S, 2011MNRAS.418..145J} are entirely consistent with the results presented here, our conclusions differ markedly in that we find substantial gains can be achieved from a local transformation of the convergence field even in the presence of realistic levels of shape noise. The source of this difference stems from performing a combined analysis with the original power spectrum, which holds complementary information, rather than considering each power spectrum in isolation. 

\subsection{Full likelihood model} \label{sec:boxcox}
The posterior in the parameters $\om$ and $\sigma_8$ as produced by contemporary cosmic shear surveys is not well approximated by a multivariate Gaussian. In order to assess whether our conclusions are affected by the limitations of the Fisher matrix formalism, we consider the case of a full mock likelihood analysis following the methodology presented in \citet{2011MNRAS.418..145J}. In this work analytical models for power spectra of convergence fields transformed via Box-Cox transformations, which include the logarithm, were derived using a perturbative expansion and recalibration of higher-order contributions via the third to fifth moments of the convergence measured from N-body simulations. \citet{2011MNRAS.418..145J} used a suite of 100 convergence fields at a single source redshift $z=1$ simulated with the SUNGLASS pipeline \citep{2011Kiessling} for signal and covariance estimation via Eq.~(\ref{eq:rescale}). We adopt these measurements for convergence field without shape noise and smoothed with a Gaussian of width 1.5 arcmin. Moreover, we rescale the covariance to correspond to a 200 $\rm{deg}^2$ survey. This toy model setup is deliberately chosen to yield wide contours with a strong non-linear degeneracy, which is least well represented by a Gaussian approximation to the posterior. We use the value for the free parameter in the logarithmic transformation, equivalent to $\kappa_0$ in Eq. (5), as determined by \citet{2011MNRAS.418..145J} in the Box-Cox optimisation.

The resulting parameter constraints for a power spectrum analysis of the original convergence fields, of the log-transformed fields, and their combined analysis are shown in Figure \ref{fig:boxcox}. We have used 10 angular frequency bins logarithmically spaced in the range 150 to 1500. The contours for the log-transformed fields are slightly biased high, which is caused by the difficulty of analytically modelling the transformed power spectrum at high $\ell$ where higher-order correlations become important (see \citealt{2011MNRAS.418..145J} for a more detailed discussion). However, this does not affect our conclusion that there is excellent qualitative agreement between the full likelihood analysis and the Fisher forecasts. The dotted contours represent a joint analysis in which the cross-covariance has been set to zero, as would arise if the two spectra were taken from fields in different parts of the sky. This greatly underestimates the achievable constraints, demonstrating the importance of the cross-correlations in suppressing noise contributions.

This independent analysis offers striking confirmation of the considerable gains which are achievable by accounting for the strong cross-covariance which exists between the power spectra of a locally transformed field and its original form. In terms of the areas enclosed by the contours in Figure \ref{fig:boxcox}, the combined analysis offers a gain in performance of more than a factor of five compared to the transformed field alone.

\section{Conclusions} \label{sec:conclusions}

In this work we have demonstrated how applying a transformation such as clipping to a convergence field can not only introduce new information to the resulting power spectrum, but that this new information helps lift the degeneracy in the $\omseight$ plane. When considering moderately large angular scales $\ell < 1500$, parameter constraints can be improved by more than a factor of two, even when realistic levels of shape noise are included. The loss of performance reported in previous studies for a single transformed power spectrum is largely attributed to the elongation of the contours along the degeneracy direction, but this elongation is readily truncated by including the original power spectrum which exhibits a different degeneracy direction. 

A second major advantage of performing a local transformation such as clipping is that it may also permit smaller angular scales to be studied, since the resulting power spectrum is more straightforward to model. We may therefore have significantly underestimated the gains achievable when clipping convergence fields. The use of tomography - dividing the source galaxies into discrete redshift bins - in conjunction with clipping will further enhance the level of precision attainable, but we leave this topic for the subject of future work. A suite of N-body cosmological simulations would be required to adequately model the clipped convergence power spectra. Many simulations will in any case be required for modelling the conventional power spectrum, or alternatively rescaling methods could prove effective for exploring a greater volume of parameter space \citep{2010AnguloWhite, 2014MeadPeacock}.  

As found when applying clipping to large scale structure, the choice of threshold is fairly flexible. Provided the threshold is sufficiently strong to reduce the large scale amplitude of the power spectrum by over $30\%$, there is little change in performance. The optimal choice ultimately rests upon the level of noise within the particular field under investigation. For the convergence fields considered in this work we find that in order for clipping to reduce the large scale power by a  factor of two, a threshold value of $\kappa_0 = 0.026$ is required, corresponding to approximately $20\%$ of the field being clipped.

The greatest obstacle to applying a local transformation to weak lensing data is that the observable field is not a direct tracer of the matter density, so we must convert the shear field to a convergence field in order to find a suitable proxy for the projected matter density. The conversion process is particularly challenging in the presence of noise and masks, but  significant progress is being made \citep{2013MNRAS.433.3373V, 2015arXiv150501871C}. We therefore envision that a combined analysis of clipped and unclipped lensing observations will help extract more robust and precise information from future and forthcoming surveys. 

\section*{Acknowledgements}
FS acknowledges support by the European Research Council under the European Community's Seventh Framework Programme FP7-IDEAS-Phys.LSS 240117.  JHD is supported by the NSERC of Canada. LV acknowledges support from the European Research Council (grant FP7-IDEAS-Phys.LSS 240117) and a Mineco grant FPA2011- 29678-C02-02. RJ acknowledges support from Mineco grant FPA2011- 29678-C02-02. CH acknowledges support from the European Research Council under the EC FP7 grant number 240185. BJ acknowledges support by an STFC Ernest Rutherford Fellowship, grant reference ST/J004421/1.

Computations for the $N$-body simulations were performed
on the GPC supercomputer at the SciNet HPC Consortium.
SciNet is funded by: the Canada Foundation for Innovation
under the auspices of Compute Canada; the Government
of Ontario; Ontario Research Fund - Research Excellence;
and the University of Toronto.

\setlength{\bibhang}{2.0em}
\setlength{\labelwidth}{0.0em}
\bibliographystyle{mn2e}   
\bibliography{../../../HomeSpace/Routines/dis} 
 
\end{document}